\documentclass[reqno,11pt]{amsart}
\usepackage{amssymb, amsmath}
\usepackage{graphicx}
\usepackage{cite}

\usepackage{subfig, tikz}
\usetikzlibrary{decorations.pathmorphing}
\usepackage{graphics}

\newcommand{\C}{\mathbb{C}}
\newcommand{\CP}{\mathbb{CP}}

\newcommand{\R}{\mathbb{R}}

\newcommand{\Z}{\mathbb{Z}}

\def\be{\begin{equation}}
\def\ee{\end{equation}}

\def\p{\partial}
\def\tw{\widetilde{W}}
\def\tz{\widetilde{Z}}

\newcommand{\hook}{{\setlength{\unitlength}{11pt}   
                   \begin{picture}(.833,.8)
                   \put(.15,.08){\line(1,0){.35}}
                   \put(.5,.08){\line(0,1){.5}}
                   \end{picture}}}

\usepackage{bm}
\usepackage{enumerate}

\newtheorem{theo}{Theorem}[section]

\newtheorem{defi}[theo]{Definition}

\newcounter{mnotecount}[section]

\renewcommand{\themnotecount}{\thesection.\arabic{mnotecount}}

\newcommand{\mnote}[1]
{\protect{\stepcounter{mnotecount}}$^{\mbox{\footnotesize
$
\bullet$\themnotecount}}$ \marginpar{
\raggedright\tiny\em
$\!\!\!\!\!\!\,\bullet$\themnotecount: #1} }

\numberwithin{equation}{section}

\begin{document}
\date{01 January 2025}
\title{Gravitational Instantons, old and new}
\author{Maciej Dunajski}
\address{Department of Applied Mathematics and Theoretical Physics\\ 
University of Cambridge\\ Wilberforce Road, Cambridge CB3 0WA\\ UK.}
\email{m.dunajski@damtp.cam.ac.uk}

\maketitle
\begin{abstract}
This is a review of gravitational instantons -  solutions to Riemannian Einstein or Einstein--Maxwell equations in four dimensions which  yield complete metrics on non--compact four--manifolds, and which asymptotically
`look like' flat space.  The review focuses on examples, and  is based on lectures given by the author at the Cracow School of Theoretical Physics held in Zakopane in June 2024.
\end{abstract}
\section{Introduction}
Gravitational instantons are solutions to the four-dimensional Einstein equations in Riemannian signature which give complete metrics and asymptotically ‘look-like’ flat space:  If $(M, g)$ is a gravitational instanton, then
\[
\int_M |\mbox{Riem}|^2\mbox{vol}_M<\infty,
\]
where $|\mbox{Riem}|^2$ is the squared $g$--norm of the Riemann tensor of $g$.

The  study of gravitational instantons
has been initiated by Stephen Hawking in his quest
for Euclidean quantum gravity \cite{H1}, and since then lot of effort has been put to make the term `look--like' into a precise mathematical statement. While  Euclidean quantum gravity  does not any-more aspire to a status of a fundamental theory, the study of gravitational instantons has influenced both theoretical physics and pure mathematics. This short review focuses on examples. It is based on lectures given by the author at the Cracow School of Theoretical Physics held in Zakopane in June 2024, and at the
Banach Center - Oberwolfach Graduate Seminar {\em Black Holes and Conformal Infinities of Spacetime}
held in Bedlewo in October 2024.
\section{Examples}
Some gravitational instantons arise as analytic continuations of Lorentzian black hole solutions to Einstein, or Einstein--Maxwell equations. If the imaginary time is turned into a periodic coordinate with the period given by the surface--gravity of Lorentzian black holes, then the resulting solutions
are regular Riemannian metrics. Euclidean Schwarzschild and Kerr metrics belong to this category. Other gravitational instantons have no Lorentzian analogues, for example because
their Riemann curvature is anti--self--dual. The Eguchi--Hanson and anti--self--dual 
Taub--NUT solutions are such examples. 
\subsection{Euclidean Schwarzschild metric}
The Schwarzschild metric is given by
\[
g=  -\left(1-\frac{2m}{r}\right)d t^2+ \left(1-\frac{2m}{r}\right)^{-1}dr^2
+r^2(d\theta^2+\sin^2{\theta}d\phi^2).
\]
The apparent  singularity at $r=2m$ corresponds
to an event horizon, and can be removed by a coordinate transformation. The singularity at $r=0$ is essential as the squared norm of the Riemann tensor blows up as $r^{-6}$. An attempt to get
rid of this singularity by removing the origin $r=0$ from the space--time leads to a geodesically incomplete
metric.

The Euclidean Schwarzschild metric \cite{H1} is obtained by setting $t=i\tau$, and restricting the range of $r$ to $2m<r<\infty$.
Set $\rho=4m\sqrt{1-2m/r}$. Near $\rho=0$ the metric takes the form
\[
 g\sim d\rho^2+\frac{\rho^2}{16m^2}d\tau^2+4m^2
(d\theta^2+\sin^2{\theta}d\phi^2).
\]
This metric is flat and regular as long as the imaginary time $\tau$ is periodic with the period $8\pi m$.  This period is inverse proportional to the Hawking temperature of the black hole radiation (Fig 1). Although this was not how the Hawking temperature was first discovered, the instanton methods gave rise
to a derivation simpler than the original calculation based on the Bogoliubov transformation \cite{GPT1, GPT2}.
In a similar manner the non--extreme Kerr black hole can be turned into the Euclidean Kerr instanton with the period of the imaginary time proportional to the inverse of the surface gravity. In the case of the extreme Kerr solution the surface gravity vanishes
and the extreme Kerr instanton does not exist.
\begin{center}
\includegraphics[scale=0.1,angle=0]{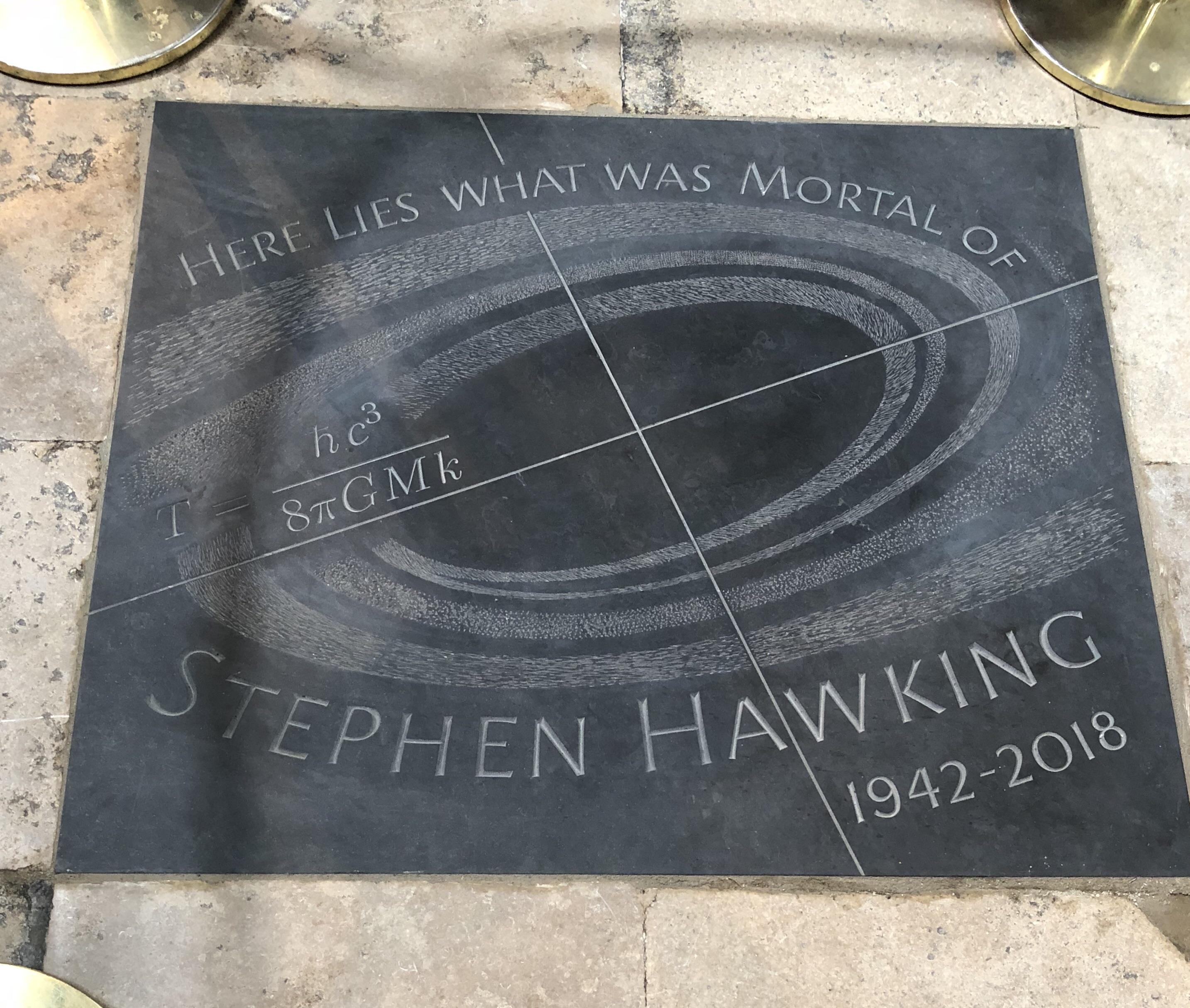}
\end{center}
\subsection{ Anti--self--dual Taub--NUT and ALF metric}
\label{sec22}
Before introducing the next example 
let us  define the left--invariant one--forms 
$(\sigma_1, \sigma_2, \sigma_3)$
on $S^3=SU(2)$ by
\[
\sigma_1+i\sigma_2=e^{-i\psi}(d\theta+i\sin{\theta} d\phi),\quad 
\sigma_3=d\psi+\cos{\theta}d\phi
\]
where $ 
0\leq\theta\leq\pi,  0\leq\phi\leq 2\pi,
0\leq\psi\leq 4\pi.$ They satisfy
\[
d\sigma_1+\sigma_2\wedge\sigma_3=0, \quad
d\sigma_2+\sigma_3\wedge\sigma_1=0,\quad d\sigma_3+\sigma_1\wedge\sigma_2=0.
\]
In terms of these one--forms the flat metric on $\R^4$ is given by
\be
\label{R4flat}
g_{\R^4}=dr^2+\frac{1}{4}r^2\Big({\sigma_{1}}^2+{\sigma_{2}}^2+{\sigma_{3}}^2\Big).
\ee
The  Taub--NUT instanton \cite{H1}
is  
\be
\label{TB}
g_{TN}=\frac{1}{4}\frac{r+m}{r-m}dr^2+m^2\frac{r-m}{r+m}\sigma_3^2+\frac{1}{4}(r^2-m^2)\left(\sigma_1^2+\sigma_2^2\right).
\ee
Introducing a coordinate $\rho$ by $r=m+\frac{\rho^2}{2m}$ shows that, near $r=m$, the metric (\ref{TB}) approaches the flat metric ({\ref{R4flat}) and so $r=m$ is only a coordinate singularity.
The Riemann curvature of the metric (\ref{TB}) is 
anti-self--dual (ASD);  it satisfies
\be
\label{ASDr}
 R_{abcd}=-\frac{1}{2}{\varepsilon_{ab}}^{pq}R_{cdpq},
 \ee
where $\varepsilon_{abcd}=\varepsilon_{[abcd]}$
is a chosen volume--form on $M$.  The ASD condition in particular 
implies the vanishing of the Ricci--tensor. This follows from taking the trace of (\ref{ASDr}). 
It also shows that the metric (\ref{TB}) has no Lorentzian analogue, as the Riemann tensor of a metric in signature $(3, 1)$ is ASD iff the metric is flat.
For large $r$ the metric $g_{TN}$ is the $S^1$ bundle over $S^2$ with Chern number equal to $1$ - this is the Hopf fibration with the total space $S^3$.

The ASD Taub--NUT example (\ref{TB}) motivates the following definition
\begin{defi}
A complete regular four-dimensional Riemannian manifold
$(M,  g)$ which solves the Einstein equations 
is called ALF (asymptotically locally flat) if it approaches $S^1$ bundle
over $S^2$ at infinity.
\end{defi}
The asymptotic form of an ALF metric is
\[
\lim_{r\rightarrow\infty} g= (d\tau+2n\cos{\theta}d\phi)^2+dr^2+r^2(d\theta^2+\sin{\theta}^2d\phi^2),
\]
where the integer $n$ is the Chern number of the $S^1$ bundle. If the $S^1$--bundle is trivial, so that $n=0$, the ALF metric is called
asymptotically flat (AF).  Euclidean Schwarzschild and
Euclidean Kerr metrics are AF. 
According to the  Lorentzian black hole uniqueness
theorems  of Hawking, Carter, D. Robinson, and others \cite{Wald}, the Kerr family of solutions exhausts all AF solutions to the Einstein equations with $\Lambda=0$. These theorems gave rise to the  Riemannian  `black hole uniqueness' conjecture stating that the Euclidean Schwarzschild and  Kerr are the only AF gravitational instantons \cite{Lapedes}.
This conjecture is now known to be false. We shall return to it in \S\ref{CTsection}.

The ASD Taub--NUT instanton, and other ALF metrics can be uplifted to the so--called Kaluza--Klein monopoles
in $4+1$--dimension \cite{KK1, KK2} with the product metric
\[
  ds^2=-dt^2+g_{TN}.
\]  
The Kaluza--Klein reduction of $ds^2$ along the Killing vector $\p/\p\psi$ gives a monopole--type solution
to the Einstein--Maxwell-dilaton theory in $(3+1)$ dimensions.

\subsection{Eguchi--Hanson and the ALE metrics}
\label{sec23}
The  Eguchi--Hanson (EH) instanton \cite{EH78, EGH} is given by
\be
\label{EH}
g_{EH}=\left(1-\frac{a^4}{r^4}\right)^{-1}dr^2+\frac{1}{4}r^2
\left(1-\frac{a^4}{r^4}\right)\sigma_3^2+\frac{1}{4}r^2\left(\sigma_1^2+\sigma_2^2\right)
\ee
with $r>a$. Setting $
\rho^2=r^2\left[1-(a/r)^4\right]$ we find that, near
$r=a$, the metric is  given by
\[
g\sim \frac{1}{4}\left(d\rho^2+\rho^2 d\psi^2\right).
\]
This metric is regular as long as the ranges of the angles are
\[
0\leq \phi\leq 2\pi, \quad 0\leq\theta\leq\pi, \quad
0\leq\psi\leq 2\pi.
\]
Thus, although for $r\rightarrow \infty$, the Eguchi--Hanson metric approaches (\ref{R4flat}), given
the allowed range of $\psi$ this metric is not asymptotically Euclidean, but corresponds to a quotient $\R^4/\Z_2$. The Eguchi--Hanson example motivates the following
\begin{defi}
A complete regular four-dimensional Riemannian manifold
$(M, g)$ which solves the Einstein equations
is called ALE (asymptotically locally Euclidean) if it approaches $\R^4/\Gamma$
 at infinity, where $\Gamma$ is a discrete
subgroup of $SO(4)$.
\end{defi}
The anti--self--dual ALE metrics are the  best understood class
of gravitational instantons. This is due to the following 
\begin{theo}[Kronheimer \cite{K1, K2}]  For any $\Gamma$ (cyclic $A_N$, dihedral $D_N$,
dihedral,
tetrahedral, octahedral, and icosahedral) there exists
an ALE gravitational instanton. 
\end{theo}
The Eguchi--Hanson metric corresponds to the case $A_2$, where $\Gamma=\Z_2$.
It is not known \cite{Gibbons_survey, nakajima} whether there exist non self--dual or anti--self--dual ALE Ricci--flat metrics.
\section{Multi--centered metrics}
Both the Taub--NUT and the Eguchi--Hanson metrics belong to the class of the so--called multi--centred
gravitational instantons. These instantons arise as superpositions of fundamental solutions to the Laplace equation on $\R^3$ via the Gibbons--Hawking ansatz \cite{GHref}. The verification of the Ricci--flat condition for this ansatz, as well as its geometric characterisation is best achieved by using an equivalent formulation of ASD Riemannian condition in terms of the hyper--K\"ahler structure. We shall
give the necessary definitions, and review the terminology in the next subsection. A more detailed discussion can be found in \cite{D}.
\subsection{Mathematical detour: Hyper--K\"ahler metrics}
We shall start with a definition
\begin{defi}
An almost complex structure on a $4$--manifold $M$
is an endomorphism $
I:TM\rightarrow TM$ such that $I^2=-\mbox{Id}$.
\end{defi}
The almost complex structure gives rise to a decomposition
\[
\C\otimes TM= T^{1, 0} M\oplus T^{0, 1} M, \quad\mbox{given by}\quad
X=\frac{1}{2}[X-iI(X)]+\frac{1}{2}[X+iI(X)]
\]
of the complexified tangent bundle into 
eigen-spaces of $I$ with eigenvalues $\pm i$.
One says that $I$ is a complex structure iff
these eigenspaces are integrable in the sense of the
Frobenius theorem, i. e.
\be
\label{NNth}
[T^{1, 0}M, T^{1, 0}M]\subset T^{1, 0} M.
\ee
A theorem of Newlander and Nirenberg justifies the terminology: $I$ is a complex structure iff there 
exists a holomorphic atlas so that
$M$ is a two--dimensional complex manifold. For example, if $M=\R^4$ and
\[
I=\begin{pmatrix}
0 & 0 & 1 & 0   \\
0 & 0 & 0 & 1\\
-1 &0 &0 &0\\
0 & -1 &0 &0
\end{pmatrix}
\]
then (\ref{NNth}) holds and the complex atlas on $M=\C^2$ consists of complex
coordinates $z_1=x_1+ix_3, z_2=x_2+ix_4$ and $T^{1, 0}M=\mbox{span}\{\p/\p z_1, \p/\p z_2 \}$.

We shall now assume that $(M, g)$ is a Riemannian four--manifold  with almost--complex
structure $I$. We say that the metric $g$ is
\begin{itemize} 
\item Hermitian if $g(X, Y)=g(IX, IY)$.
\item K\"ahler if $I$ is a complex structure, and $d\Omega=0$, where $\Omega(X, Y)=g(X, IY)$. 
\item hyper--K\"ahler if it is K\"ahler
w.r.t. three complex structures $I_1, I_2, I_3$ such that
\[
I_1I_2=I_3, \quad I_2 I_3=I_1, \quad
I_3I_1=I_2.
\]
\end{itemize}
For example, if $M=\R^4$ then the metric $g=|dz_1|^2+|dz_2|^2$ is hyper--K\"ahler with
\[
\Omega_1=\frac{i}{2}(dz_1\wedge d\bar{z}_1+dz_2\wedge d\bar{z}_2), \quad \Omega_2+i\Omega_3=dz_1\wedge dz_2.
\]
The importance of  hyper--K\"ahler metrics in the study of gravitational instantons comes from the fact that 
locally, and with the choice of orientation which makes the K\"ahler forms self--dual (SD),
the Riemann tensor of $(M, g)$ anti--self--dual (ASD) iff $(M, g)$ hyper--K\"ahler. Therefore the
ASD gravitational instantons are complete
hyper--K\"ahler metrics. Compact hyper--K\"ahler
metrics are far more rare. There is the four--dimensional torus with a flat metric, and the elusive  $K3$ surface
whose existence follows from Yau's proof \cite{Yau} of the Calabi conjecture. Finding the explicit closed form of a metric on a $K3$ surface is one of the biggest open problems in the field.
\subsection{Gibbons--Hawking ansatz}
Let $(V, A)$ be respectively a function, and a one--form on $\R^3$. The metric
\be
\label{GH}
g={V}(d {x_1}^2+d {x_2}^2+d {x_3}^2)+{V}^{-1}(d \tau+A)^2,
\ee
is  hyper--K\"ahler (and therefore ASD and Ricci flat)
with the K\"ahler forms given by
\[
\Omega_i=-(d\tau+A)\wedge dx_i+\frac{1}{2}V\epsilon_{ijk} dx_j\wedge dx_k, \quad i=1, 2, 3
\]
iff the Abelian Monopole Equation
\be
\label{monopole3}
dA=\star_3 dV 
\ee
holds 
(here $\star_k$ is the Hodge operator on $\R^k$ taken w.r.t the flat metric and a chosen volume form). This equation follows from the closure condition $d\Omega_i=0$, and implies that the function $V$
is harmonic on $\R^3$. 
The general Gibbons Hawking ansatz (\ref{GH}) is characterised by the hyper--K\"ahler condition together with the
existence of a Killing vector $K$ which Lie--derives all K\"ahler forms. The Cartesian coordinates $(x_1, x_2, x_3)$
in (\ref{GH}) arise as the moment maps, i. e. $K\hook\Omega_i=dx_i$.

The multi--centre metrics
correspond to a choice
\be
\label{formV}
V = V_0 + \sum_{m=1}^N \frac{1}{\mid {\bf x} - {\bf x}_m \mid},
\ee
where $V_0$ is a constant, and ${\bf x}_1, \dots,
{\bf x}_N$ are position vectors of $N$ points in $\R^3$. The special cases of (\ref{formV}) are
\begin{itemize}
\item $V_0=0, N=1$ give  the flat metric.
\item $V_0=0, N=2$ give the  Eguchi--Hanson metric (\ref{EH}) albeit in a different coordinate system.
$V_0=0$ and  $N>2$ correspond to the general $A_{N}$ ALE instantons.
\item $V_0\neq 0, N=1$ give the Taub--NUT metric (\ref{TB}). $V_0\neq 0, N>1$ correspond to the
$A_{N}$ ALF instantons.
\end{itemize}
\section{The Chen--Teo instanton}
\label{CTsection}
The  Riemannian 
black hole uniqueness conjecture we alluded to in \S\ref{sec22} is now known to be wrong. Chen and Teo \cite{CT1, CT2}
have constructed a five parameter family of toric (i. e. admitting two commuting Killing vectors) Riemannian Ricci flat metrics interpolating between the 
ALE three--centre Gibbons--Hawking metrics with centres on one axis,  and
Euclidean Pleba\'nski--Demia\'nski solutions \cite{PD}. The Chen--Teo family  contains a two--parameter sub--family
of AF instantons which are not in the Euclidean Kerr family of solutions. It has been proven by
Aksteiner and Andersson \cite{AA} that, as the Chen--Teo family consists of Hermitian and therefore one--sided Petrov--Penrose type D solutions, the Chen--Teo instantons do not arise as an an analytic continuation of any Lorentzian black holes.
\subsection{Explicit formulae}
Let $f$ be a quartic polynomial with four  real roots. Set
\begin{eqnarray*}
f&=&f(\xi)=a_4\xi^4+a_3 \xi^3+a_2\xi^2+a_1\xi+a_0\\
F&=&f(x)y^2-f(y)x^2\\
H&=& (\nu x+y)[(\nu x-y)(a_1-a_3xy)-2(1-\nu)(a_0-a_4 x^2y^2)]\nonumber\\
G&=& f(x)[(2\nu-1)a_4 y^4+2\nu a_3  y^3+a_0 \nu^2]-f(y)[\nu^2 a_4 x^4+2\nu a_1 x+(2\nu-1)a_0].\nonumber
\end{eqnarray*}
The family of metrics
\be
\label{CT}
g=\frac{kH}{(x-y)^3}\Big(\frac{dx^2}{f(x)}-\frac{dy^2}{f(y)}-\frac{f(x)f(y)}{kF} d\phi^2\Big)
+\frac{1}{FH(x-y)}(F d\tau+Gd\phi)^2
\ee
is Ricci--flat for any choice of the parameters $(a_0, \dots, a_4, \nu, k)$.
Two out of five parameters $(a_0, \dots, a_4)$ can be fixed by scalings, so (\ref{CT}) is a five--parameter family. The  Riemann curvature is regular
if the range of $(x, y)$ is restricted to the rectangle
on Figure 2, where $r_1<r_2<r_3<r_4$ are the roots  $f$.
\begin{center}
\includegraphics[scale=0.25,angle=0]{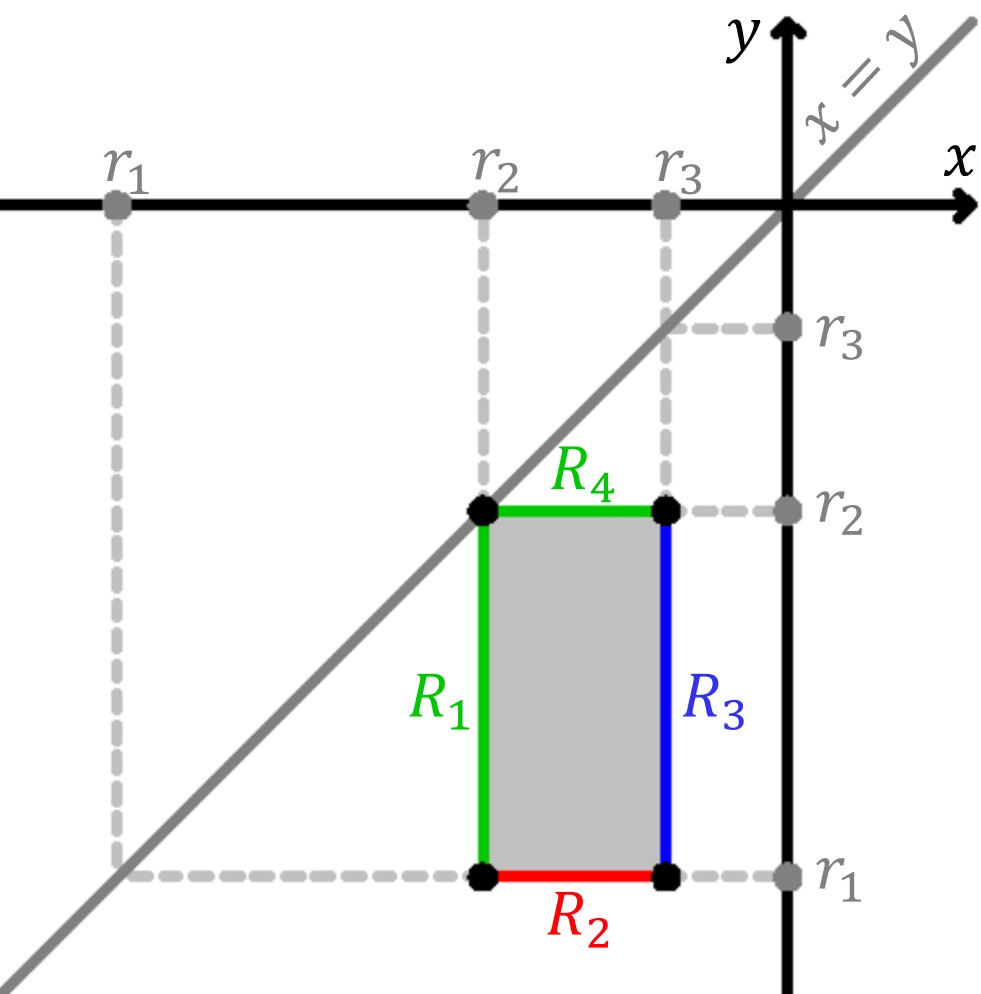}
\end{center}
To avoid the conical singularities, and ensure the asymptotic flatness one makes a choice
\be
\label{ctroots}
r_1=\frac{4s^2(1-s)}{1-2s+2s^2}, \; r_2=-1, \; r_3=\frac{1-2s}{s(1-2s+2s^2) } , \; r_4=\infty,\;
\nu=-2 s^2, \quad s\in(1/2, \sqrt{2}/2).
\ee
This leads to a two parameter family of AF instantons on $M=\CP^2\setminus S^1$. 
\subsection{The rod structure}
The Chen--Teo metrics  (\ref{CT}) admit two commuting Killing vectors  $K_i=\p/\p\phi^i$ where $\phi^i=(\phi, \tau)$. Any metric with two commuting Killing vectors can locally be put in the form 
\be
\label{toric}
g=\Omega^2(dr^2+dz^2) +J_{ij} d\phi^id\phi^j, \quad i, j=1, 2
\ee
where $J=J(r, z)$ is a $2$ by $2$ symmetric matrix, and the $(r, z)$ coordinates are defined by
\[
  r^2=\mbox{det}(J), \quad \star_2 dz = dr.
\]
The space of orbits of the $T^2$ action is the upper half--plane $\mathbb{H}=\{(r, z), r>0\}$ with the boundary $\partial \mathbb{H}$ where  rank$(J(0, z))<2$. Generically this rank is equal to $1$. It vanishes at the turning points $z_1, z_2, \dots, z_N$
where both Killing vectors vanish. These turning points divide the $z$--axis into $(N+1)$ rods \cite{Har} 
\[
I_1=(-\infty, z_1),  I_2=(z_1, z_2), \dots, I_N= (z_{N-1}, z_N), I_{\infty}=(z_N, \infty).
\]
In the Lorentzian case these rods correspond to horizons or axes of rotation, and in the Riemannian case they are axes. The rod data associated to (\ref{toric})  consists of a collection of $(N+1)$ rods, together with the lengths $(z_k-z_{k-1}), k=2, \dots, N$ of the finite rods, and the constant rod vectors $V_2, \dots, V_N$ such that $V_k$ vanishes on the rod $I_k$. Each of these vectors can be expanded as
$V_k=V_k^1 K_1+V_k^2 K_2$, and then the admissibility condition \cite{Hollands} is 
\[
\det\begin{pmatrix}
V_k^1 & V_k^2 \\
V_{k+1}^1 & V_{k+1}^2
\end{pmatrix}=\pm 1.
\]

 While the rod structure does not uniquely determines the metric of the instanton, it specifies the 
topology of the underlying four--manifold \cite{Nilsson}. The number of turning points is equal to the Euler signature.  In the Chen--Teo case there exist thee turning points, 
so that $\chi(M)=3$ for the Chen--Teo instanton. Closing up the semi--infinite rods gives the triangular rod structure
of $\CP^2$ with three turning points as the triangle vertices, and three rods as sides. Joining the rods adds $S^1\times \R^3$ to $M$, and so $M=\CP^1\setminus{S^1\times \R^3}\cong \CP^1\setminus{S^1}$.
The signature of the Chen--Teo family is  $1$.
\subsection{The Yang equation and ASDYM}
The Ricci--flat condition on (\ref{toric}) reduces to the Yang equation
\be
\label{Yeq}
r^{-1} \p_r (r J^{-1}\p_r J)+\p_z(J^{-1}\p_z J)=0.
\ee
Once a solution to this equation has been found, the conformal factor $\Omega$ can be found by a single integration.

The Yang  equation (\ref{Yeq}) also arises as a reduction of anti--self--dual Yang-Mills equations \cite{Witten, W2}. To see it,  consider the complexified Minkowski space $M_{\C}=\C^4$, with coordinates
$(W, Z, \tw, \tz)$ such that the metric and the volume form are 
\[
ds^2=2(dZ d\tz-dWd\tw), \quad \mbox{vol}=dW\wedge d\tw \wedge dZ \wedge d\tz.
\]
Let $\Phi\in\Lambda^1(M_{\C})\otimes\mathfrak{sl}(2)$, and $ F=d\Phi+\Phi\wedge \Phi$. The anti--self--dual Yang--Mills (ASDYM)
equations are $F=-\star_4 F$ (now $\star_4$ is taken w.r.t. the flat metric on $\C^4$), or  
\be
\label{ASDYM}
F_{WZ}=0, \quad F_{\tw\tz}=0,\quad  F_{W\tw}-F_{Z\tz}=0.
\ee
The first two equations imply the existence of a gauge choice such that 
\be
\label{eqPhi}
\Phi=J^{-1}\p_{\tw} J d\tw+J^{-1}\p_{\tz} J d\tz, \quad J=J(W, Z, \tw, \tz)\in SL(2, \C).
\ee
The final equation in (\ref{ASDYM}) holds iff
\be
\label{ASDYMJ}
\p_Z(J^{-1}\p_{\tz} J)-  \p_W(J^{-1}\p_{\tw} J)=0.
\ee
Setting
\[
Z=t+z, \quad \tz=t-z, \quad W=re^{i\theta}, \quad
\tw=re^{-i\theta},
\]
and performing a symmetry reduction $J=J(r, z)$ reduces (\ref{ASDYMJ}) to (\ref{Yeq}).
\subsection {Twistor construction}
The twistor correspondence for ASDYM is based on an observation that
ASDYM condition is equivalent to the flatness of a 
connection $\Phi$  on $\alpha$--planes in $M_{\C}$
\be
\label{incidence}
\mu=W+\lambda\tz, \quad \nu= Z+\lambda\tw.
\ee
The  twistor space $PT\equiv \CP^3\setminus\CP^1$ is the space of all such planes.
It can be covered by two open sets, with  affine coordinates $(\mu, \nu, \lambda)$ in an open set
where $\lambda \neq \infty$. Points in $M_{\C}$ correspond to rational curves 
(twistor lines) in $PT$, and points in $PT$ correspond to $\alpha$--planes in $M_{\C}$. The conformal structure
on ${M_{\C}}$ is encoded in the algebraic geometry of curves in $PT$: $p_1, p_2$ are null separated iff $L_1, L_2$ intersect. 

The connection between twistor theory and ASDYM is provided by the following 
\begin{theo}[Ward \cite{W1}]
\label{theoWard}
There exists a $1-1$ correspondence between
gauge equivalence classes of ASDYM connections $\Phi$, and holomorphic vector bundles $E\rightarrow PT$ trivial on twistor lines.
\end{theo}
To read off the solution (\ref{ASDYMJ}) from this Theorem
cover $PT$ with two open sets: $U$, where $\lambda\neq \infty$ and $\widetilde{U}$ where $\lambda\neq 0$.
The bundle $E$ is then characterised by its
patching matrix: $P=P(\mu, \nu, \lambda)$.
The triviality on twistor lines implies that there exists a splitting
$P=P_U{{P}_{\widetilde{U}}}^{-1}$, where $P_U$ and ${P}_{\widetilde{U}}$
are holomorphic and invertible matrices on $U$ and 
$\widetilde{U}$ respectively. The incidence relation (\ref{incidence}) implies that  $P$ is constant along
the vector fields $\{\p_{\tz}-\lambda\p_W, \p_{\tw}-\lambda\p_Z\}$. Applying this to the splitting relation, and
using the Liouville theorem implies the existence of $\Phi \in\Lambda^1(M_{\C})\otimes\mathfrak{sl}(2)  $ such that 
\[
\Phi=\widetilde{H}^{-1}\p_Z \widetilde{H}\; dZ+\widetilde{H}^{-1}\p_W \widetilde{H}\;dW+{H}^{-1}\p_{\tz} {H}\; d\tz+
{H}^{-1}\p_{\tw} {H}\; d\tw
\]
where $H=P_U(\lambda=0), \widetilde{H}=P_{\widetilde{U}}(\lambda=\infty)$. This is gauge equivalent to
(\ref{eqPhi}) with 
\be
\label{Jfinal}
J=H\widetilde{H}^{-1}.
\ee
\subsection{Twistor bundle for toric Ricci flat metrics}
 Let us  go back to the toric Ricci--flat metrics. For any of the Killing vectors $K$ we can find its twist potential: a function
  $\psi$ such that
  \[
  \label{twist_pot}
    d\psi =* (K\wedge dK).
  \]
  Another solution to the Yang equation 
  (\ref{Yeq})  then arises from a B\"acklund transformation
  \[
  J'=\frac{1}{V}
\begin{pmatrix}
1 & -\psi \\
-\psi & \psi^2-V^{2}
\end{pmatrix}, \quad V\equiv g(K, K).
\]
Pick a rod on which $K$ is not identically zero. The following has been established in \cite{FW, WM, MWbook}:
The patching matrix for the bundle $E$ from Theorem \ref{theoWard} is an analytic continuation of
$P(z)\equiv J'(0, z)$:
\[
P(\gamma), \quad\mbox{where}\quad \gamma=z+\frac{1}{2}r\Big(\lambda-\frac{1}{\lambda}\Big).
\]
The splitting procedure leads, via (\ref{Jfinal}), to $J'(r, z)$ from which $J(r, z)$ can be recovered.

This patching matrix can be found for the Chen--Teo family \cite{DT24}. It is given by
\be
\label{PCT}
P(z)=
\begin{pmatrix}
C_1/C & Q/C   \\
Q/C & C_2/C
\end{pmatrix},
\ee
where 
$C_1, C_2, C$ monic cubics, $Q$ quadratic, with coefficients depending on the Chen--Teo parameters.
Examining the outer rod and the asymptotics near $z=\infty$  gives
\[
P\cong
\begin{pmatrix}
1 & 0   \\
0 & -1
\end{pmatrix}+
\frac{1}{z}\begin{pmatrix}
2m & 2n   \\
2n & 2m
\end{pmatrix}
+O(1/z^2),
\]
where $m$, $n$ are mass and nut parameters.
For Chen--Teo instanton with (\ref{ctroots}) we find
\[
m=\sqrt{k}\frac{(1+2s^2)^2}{2\sqrt{1-4s^4}} ,\quad n=0
\]
in agreement with \cite{KL21}. In general, the patching matrix $P$ of the form (\ref{PCT}) where
$C, C_1, C_2$ are monic polynomials of degree $N$ and $Q$ is a polynomial of degree $N-1$ subject
to $\mbox{det}(P)=-1$ lead to Ricci--flat ALF metrics with $N+1$ rods and $N$  turning points. The ALE
metrics with $N+1$ rods can also be constructed, but from a different ansatz \cite{T25, DMT25}.
\section{Other developments}
\subsection{ALF,  ALE,  ALG,  ALH,  and more}
The ALE and ALF classes of gravitational instantons have been defined in
(\ref{sec22}) and (\ref{sec23}) in terms of the asymptotic quotients of $\R^4$  and asymptotic
$S^2$ fibrations respectively. There is an alternative and unifying definition in terms of the
volume growth of a ball of large radius $R$. It is of orders $R^4$ and $R^3$ for respectively
ALE and ALF.  This classification gives rise to 
more families of instantons: ALG and ALG*  the volume growth $R^2$, 
ALH with the volume growth $R$, and ALH* with the volume growth $R^{4/3}$ \cite{BM_paper, Hein, CCpaper}.  Unlike the ALE and ALF, these new families do not contain any examples which are known analytically in closed form.
It is however the case that all  classes are asymptotically described by the Gibbons--Hawking form 
(\ref{GH}) with the harmonic function given by
\begin{eqnarray*}
V\sim \frac{N}{|{\bf x}|} &&\quad\mbox{for ALE}\\
V\sim 1+ \frac{N}{|{\bf x}|} &&\quad\mbox{for ALF}\\
V\sim 1+\frac{N}{2\pi}\ln{({x_1}^2+{x_2}^2)} && \quad\mbox{for ALG and ALG*}\\
V\sim 1+N x_3  && \quad\mbox{for ALH and ALH*.}
\end{eqnarray*}
Therefore the metrics are locally asymptotic to
$\R^k\times T^{4-k}$ with $k=4$ for ALE, $k=3$
for ALF, $k=2$ for ALG and $k=1$ for ALH. Let us
focus on the ALH* case, and perform an affine transformation of $x_3$, such that $V=x_3$
in the Gibbons--Hawking ansatz (\ref{GH}). The coordinate $x_3$ is on the base $\R$
of the fibration $M\rightarrow \R$. The fibres
are Nil 3--manifolds fibering over $T^2$ with periodic coordinates $(x_1, x_2)$ with the fibre coordinate $\tau$. The one--form $A$ in the ansatz (\ref{GH})
is such that $dA=dx_1\wedge dx_2$ is the volume form on $T^2$. Setting $x_3=r^{2/3}$ and rescalling $(x_1, x_2, \tau)$ by constants 
yields
\[
g=dr^2+r^{2/3}(dx_1^2+dx_2^2)+r^{-2/3}(d\tau
+A)^2.
\]
The volume form is 
$\mbox{vol}=r^{1/3}dr\wedge dx_1\wedge dx_2\wedge d\tau$, so that the volume growth is
indeed $\int_M \mbox{vol}\sim R^{4/3}$ if the range of $r$ is bounded by $R$.
\subsection{Einstein--Maxwell instantons}
The gravitational instantons exist in the Einstein--Maxwell theory. Unlike the pure Einstein case, there
exist many asymptotically flat solutions in the multi--centred class. These solutions arise as analytic continuations of the Israel--Wilson and  Majumdar--Papapetrou black holes (see \cite{Whitt:1984wk,
Yulle, DH07}), and are given by 
\be
\label{IWP}
g=V\widetilde{V}(dx_1^2+dx_2^2+dx_3^2)+\frac{1}{V\widetilde{V}}(d\tau+A)^2
\ee
where $V$ and $\widetilde{V}$ are harmonic functions on $\R^3$, and the one--form $A$ satisfies
\be
\label{IWPmono}
\star_3(\widetilde{V}dV-Vd\widetilde{V})=dA.
\ee
The Maxwell field is given by
\[
F=\p_i (V^{-1}-\widetilde{V}^{-1})(d\tau+A)\wedge dx^i+\epsilon_{ijk}\p_k
(V^{-1}+\widetilde{V}^{-1}) V\widetilde{V} dx^i\wedge dx^j.
\]
If $\tilde{V}=1$ then (\ref{IWPmono}) reduces to the monopole equation (\ref{monopole3}) and the metrics (\ref{IWP}) are Ricci flat, and coincide with the Gibbons--Hawking ansatz (\ref{GH}). If
\[
V=V_0+\sum_{m=1}^N\frac{a_m}{|{\bf x}-{\bf x}_m|}, \quad 
\widetilde{V}=\widetilde{V}_0+\sum_{m=1}^N\frac{\tilde{a}_m}{|{\bf x}-{\bf \tilde{x}}_m|}
\]
with $V_0, \widetilde{V}_0, a_m, \tilde{a}_m, {\bf x}_m,
{\bf \tilde{x}}_m$ constant and $N, \widetilde{N}$
integers. In particular if $V_0=\widetilde{V}_0\neq 0, N=\widetilde{N}$ and $\sum a_m=\sum \tilde{a}_m$ then the metrics (\ref{IWP}) are AF. The Riemannian Majumdar--Papapetrou metrics have $V=\widetilde{V}$ and purely magnetic Maxwell field $F=-2\star_3 d V$. See \cite{DH07} for other choices which lead to AE, ALE and ALF solutions.

There also exist
Einstein--Maxwell instantons with no Lorentzian counterpart, and anti--self--dual Weyl 
curvature \cite{lebrun, lebrun2}. An example is the Burns metric 
\be
\label{burns}
g_{\mbox{Burns}}=dr^2+\frac{1}{4}r^2 \Big({\sigma_{1}}^2+{\sigma_{2}}^2+{\sigma_{3}}^2\Big)
+\frac{m}{4}({\sigma_1}^2+{\sigma_2}^2).
\ee
It is  the unique scalar–flat K{\"a}hler metric on the total space of the line
bundle $\mathcal{O}(-1)\rightarrow \CP^1$. It is also  an AE Einstein–Maxwell gravitational instanton, with the self–dual part of the Maxwell field strength given by the K\"ahler form, and its anti--self--dual part given by the Ricci form. It is one of few
gravitational instantons where the isometric embedding class is known: It has been shown in \cite{DTembeddings} that (\ref{burns}) can be isometrically embedded in $\R^7$, but not in $\R^6$.
\subsection{Twistor Theory and non--linear graviton}
The twistor non--linear graviton approach of Penrose \cite{Pe76} parametrises holomorphic
anti--self--dual Ricci flat metrics in terms of complex three-folds with 4--parameter family of rational curves
and some additional structures. The Riemannian version of this correspondence have been given by
Atiyah, Hitchin and Singer \cite{AHS78}, where the twistor space is the six--dimensional manifold
arising as an $S^2$--bundle over a Riemannian manifold $(M, g)$. Each fiber of the $S^2$--fibration parametrises the almost--complex structures in $M$.
The twistor space is itself an almost--complex manifold, and its almost--complex structure is integrable iff 
(with respect to a chosen orientation on $M$) the Weyl tensor of $g$ is ASD.
\begin{theo}[\cite{Pe76}, \cite{AHS78}]
Hyper--K{\"a}hler four--manifolds (ASD Ricci flat metrics) are 
in one-to-one correspondence with three dimensional complex manifolds (twistor spaces) admitting
4-parameter families of rational curves with some additional structure.
\end{theo}
This formulation is well suited to the study of gravitational instantons. In particular
the ALE class can be fully characterised twistorially \cite{Hitchin, K1, K2, Hitchin2}. In this case there exists a holomorphic
fibration $PT\rightarrow {\mathcal O}(k)$ for some integer $k$. If $k=2$, then the associated instanton
admits a tri--holomorphic Killing vector and belongs to the $A_N$ Gibbons--Hawking class (\ref{GH}), \cite{TW}. If $k>2$ then in general $(M, g)$ does not admit a Killing vector, but it admits tri--holomorphic Killing spinor 
which leads to a hidden symmetry of the associated heavenly equations \cite{DM1, DM2}.
\subsection{Euclidean quantum gravity}
Euclidean quantum gravity which gave rise to the initial interest in gravitational instantons in the late 1970 does
not any more aspire to the status of a fundamental theory of quantum gravity. According to Gary Gibbons's interesting account \cite{Garry_book}, it never did. And yet it is the only theory of quantum gravity with experimental predictions,
including the black hole thermodynamics. In this theory the gravitational instantons dominate the Euclidean path integral. So if a quantum gravity theory exists, and if it reduces to Einstein's
general relativity in the classical limit, then Euclidean quantum gravity is here to stay, and will occupy a place
similar to that the WKB approximation has in the quasi-classical limit relating the quantum mechanics to
Newtonian physics. This short, and subjective review has focused on recent, and not so recent,
mathematical development. It remains to be seen what role will the gravitational instantons play in physics in the years to come.

\end{document}